\documentclass[nofootinbib,twocolumn,showpacs,preprintnumbers,pre,aps]{revtex4-1}

\usepackage[dvipdfmx]{graphicx}
\usepackage{bm}
\usepackage{amsmath}
\usepackage{amssymb}
\usepackage{color}
\usepackage{here}
\usepackage{siunitx}

\begin{document}

\title{Hydrodynamic Interaction between Two Elastic Microswimmers}

\author{Mizuki Kuroda}

\author{Kento Yasuda}

\author{Shigeyuki Komura}
\email{komura@tmu.ac.jp}

\affiliation{
Department of Chemistry, Graduate School of Science,
Tokyo Metropolitan University, Tokyo 192-0397, Japan}


\begin{abstract}
We investigate the hydrodynamic interaction between two elastic swimmers composed of 
three spheres and two harmonic springs. 
In this model, the natural length of each spring is assumed to undergo a prescribed cyclic change, 
representing the internal states of the swimmer [K. Yasuda \textit{et al.}, J.\ Phys.\ Soc.\ Jpn.\ \textbf{86}, 
093801 (2017)].
We obtain the average velocities of two identical elastic swimmers as a function of the distance 
between them for both structurally asymmetric and symmetric swimmers.
We show that the mean velocity of the two swimmers is always smaller than that of a 
single elastic swimmer. 
The swimming state of two elastic swimmers can be either bound or unbound depending on the relative 
phase difference between them.
\end{abstract}

\maketitle

\section{Introduction}
\label{sec:introduction}

Microswimmers are small machines that swim in a fluid and they are expected to be used in microfluidics 
and microsystems~\cite{Lauga09}.
Over the length scale of microswimmers, the fluid forces acting on them are dominated by the frictional 
viscous forces.
By transforming chemical energy into mechanical energy, however, microswimmers change their 
shape and move efficiently in viscous environments.
According to Purcell's scallop theorem, time-reversible body motion cannot be 
used for locomotion in a Newtonian fluid~\cite{Purcell77,Lauga11}.
As one of the simplest models exhibiting broken time-reversal symmetry, 
Najafi and Golestanian proposed a three-sphere swimmer (NG swimmer)~\cite{Golestanian04,Golestanian08}, 
in which three in-line spheres are linked by two arms of varying length.
Recently, such a swimmer has been experimentally realized by using colloidal beads manipulated by 
optical tweezers~\cite{Leoni09}, ferromagnetic particles at an air-water 
interface~\cite{Grosjean16,Grosjean18}, or neutrally buoyant spheres in a viscous 
fluid~\cite{Box17}.

Using the NG swimmer model, Pooley \textit{et al.}\ showed that the interaction 
between two swimmers depends on their relative displacement, orientation, and 
phase, leading to motion that can be either attractive, repulsive, or oscillatory~\cite{Pooley07}. 
Scattering of two NG swimmers was also investigated on the basis of the time-reversal 
invariance of the Stokes equation~\cite{Alexander08}.
Later Farzin \textit{et al.}\ reexamined the hydrodynamic interaction between two NG swimmers
and concluded that the long-time swimming states are different between moving in the 
same and opposite directions~\cite{Farzin12}.
To understand hydrodynamic coupling for stochastic swimmers, on the other hand, Najafi and 
Golestanian studied the correlated motion of a three-sphere swimmer and a two-sphere 
system~\cite{Najafi10}.
They calculated the swimming velocities as functions of the statistical transition rates for the 
conformational changes.

Recently, the present authors have proposed a generalized three-sphere microswimmer model 
in which the spheres are connected by two harmonic springs, i.e., an elastic 
microswimmer~\cite{Yasuda17}.  
Compared with the NG swimmer, the main difference of the elastic swimmer is that the natural 
length of each spring (rather than the arm length) oscillates in time and is assumed to undergo 
a prescribed cyclic change. 
As a result, the sphere motion in our model is determined by the natural spring lengths, representing 
the internal states of a swimmer, and also by the force exerted by the fluid. 
We have analytically obtained the average swimming velocity as a function of the frequency of 
the cyclic change in the natural length~\cite{Yasuda17}.
In the low-frequency region, the swimming velocity increases with frequency and it reduces 
to that of the NG swimmer~\cite{Golestanian04,Golestanian08}.
Conversely, in the high-frequency region, the velocity decreases with increasing frequency.
We note that similar models were proposed by other people~\cite{Dunkel09,Pande15,Pande17}, 
while our elastic swimmer model was further extended to thermally driven elastic 
micromachines~\cite{Hosaka17}.

In this work, we investigate the hydrodynamic interaction between two elastic three-sphere 
swimmers that are confined in one-dimensional space and moving in the same direction.
We first derive a general expression for the average velocities (over a period of one cycle) of two 
hydrodynamically interacting three-sphere swimmers as a function of the distance between them.
Using this general expression, we then calculate the explicit forms of the average velocities of 
two identical elastic microswimmers. 
We show that the mean of the two average velocities is always smaller than that of 
a single elastic swimmer,  whereas the velocity difference depends on the relative phase difference 
in the natural lengths between the two swimmers. 
As a result, the swimming state of two elastic swimmers can be either bound or unbound 
depending on the relative phase difference.

In Sect.~\ref{sec:two}, we first discuss the motion of two interacting three-sphere microswimmers.
In Sect.~\ref{sec:elastic}, we calculate the average velocities of two interacting elastic swimmers, 
and further discuss the mean and the difference between the two average velocities. 
The average velocities of two symmetric elastic swimmers is discussed in 
Sect.~\ref{sec:symmetric}. 
In Sect.~\ref{sec:NG}, we discuss the interaction of two NG swimmers by considering 
the low-frequency limit of our results.
Finally, a summary of our work and some discussion are given in Sect.~\ref{sec:summary}.

\section{Two interacting three-sphere swimmers}
\label{sec:two}

\begin{figure}[tbh]
\centering
\includegraphics[scale=0.2]{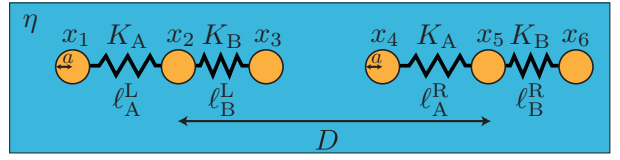}
\caption{(Color online)
Two elastic three-sphere microswimmers in a viscous fluid characterized by shear viscosity 
$\eta$. 
The positions of the three spheres in the left (L) swimmer are denoted by $x_1$, $x_2$, and $x_3$ in 
a one-dimensional coordinate system, while those in the right (R) swimmer are denoted by 
$x_4$, $x_5$, and $x_6$.
The distance between these two swimmers is defined by $D=x_5-x_2$. 
In each elastic swimmer, three identical spheres of radius $a$ are connected by two harmonic springs
A and B characterized by elastic constants $K_{\rm A}$ and $K_{\rm B}$, respectively.  
The four natural lengths of the springs, $\ell^{\rm L}_{\rm A}$, $\ell^{\rm L}_{\rm B}$, 
$\ell^{\rm R}_{\rm A}$, and $\ell^{\rm R}_{\rm B}$, depend on time and are assumed to undergo 
cyclic changes as given by Eqs.~(\ref{ellLA})--(\ref{ellRB}).
}
\label{fig:twoswimmerpic}
\end{figure}

As shown in Fig.~\ref{fig:twoswimmerpic}, we consider two general three-sphere swimmers 
in a viscous fluid characterized by shear viscosity $\eta$. 
Each swimmer consists of three hard spheres of the same radius $a$ connected either by two arms 
(NG swimmer) or by two harmonic springs (elastic swimmer explained in the next section) A and B. 
The positions of the three spheres in the left (L) swimmer are denoted by $x_1$, $x_2$, and $x_3$ 
in a one-dimensional coordinate system, while those in the right (R) swimmer are denoted by 
$x_4$, $x_5$, and $x_6$. 
We also assume $x_1<x_2<x_3 \ll x_4<x_5<x_6$ without loss of generality.
The distance between the two swimmers is defined by the positions of the middle spheres, i.e.,
$D=x_5-x_2$.

Owing to the hydrodynamic interaction, each sphere exerts a force on the viscous fluid and 
is subjected to an opposite force from it.
Denoting the velocity of each sphere by $\dot x_i=dx_i/dt$ and the force acting on each sphere by $f_i$
($i=1, \dots, 6$), we can write the equations of motion of each sphere as
\begin{equation}
\dot x_i  = \sum_{j=1}^{6}M_{ij}f_j,
\label{ssokudo}
\end{equation}
where the details of the hydrodynamic interactions are taken into account through the mobility 
coefficients $M_{ij}$.
Within Oseen's approximation, which is justified when the spheres are considerably far from each other
($a \ll \vert x_i-x_j \vert$), the expressions for the mobility coefficients $M_{ij}$ can be written as 
\begin{equation}
M_{ij}=
\begin{cases}
\dfrac{1}{6\pi\eta a}, & i=j, \\
\dfrac{1}{4\pi\eta \vert x_i-x_j \vert}, & i\neq j.
\end{cases}
\label{joken}
\end{equation}
We remark here that although the two swimmers are aligned along the one-dimensional axis, 
the spheres are interacting through the three-dimensional hydrodynamic interaction.
Furthermore, we require two force-free conditions of the two swimmers, i.e., 
\begin{align}
f_1+f_2+f_3=0, ~~~~f_4+f_5+f_6=0.
\label{forcefree}
\end{align}

Let us denote the average arm length (NG swimmer) or the average natural length (elastic swimmer)
by $\ell$, and also introduce the four displacements of the springs with respect to $\ell$
for the left and right swimmers as 
\begin{align}
& u^{\rm L}_{\rm A}=x_2-x_1-\ell,~~~~u^{\rm L}_{\rm B}=x_3-x_2-\ell,
\label{uL} \\
& u^{\rm R}_{\rm A}=x_5-x_4-\ell,~~~~u^{\rm R}_{\rm B}=x_6-x_5-\ell.
\label{uR}
\end{align}
Then the four kinematic constraints are given by taking the time derivative of the above relations:
\begin{align}
& \dot{u}^{\rm L}_{\rm A}=\dot{x}_2-\dot{x}_1,~~~~\dot{u}^{\rm L}_{\rm B}=\dot{x}_3-\dot{x}_2,
\label{kineleft}
\\
&\dot{u}^{\rm R}_{\rm A}=\dot{x}_5-\dot{x}_4,~~~~\dot{u}^{\rm R}_{\rm B}=\dot{x}_6-\dot{x}_5.
\label{kineright}
\end{align}

We note here that Eq.~(\ref{ssokudo}) implies six coupled equations. 
Together with the two force-free conditions in Eq.~(\ref{forcefree}) and the four kinematic constraints 
in Eqs.~(\ref{kineleft}) and (\ref{kineright}), we have sufficient equations to solve the 
twelve unknowns, namely, $\dot{x}_i$ and $f_i$ ($i=1, \dots, 6$).
Finally the average velocities of the left and right swimmers can be obtained by 
\begin{align}
V^{\rm L}=\frac{1}{3}\left \langle \dot{x}_1+\dot{x}_2+\dot{x}_3 \right \rangle,~~
V^{\rm R}=\frac{1}{3}\left \langle \dot{x}_4+\dot{x}_5+\dot{x}_6 \right \rangle,
\end{align}
where averaging $\langle \cdots \rangle$ is performed by time integration in a full cycle.

Under the condition that the two swimmers are far from each other and the deformations are small 
compared with the average arm length $\ell$, i.e., $a \ll u^{\rm L,R}_{\rm A,B} \ll \ell \ll D$,
one can perform a perturbative calculation to obtain the average velocities as
\begin{align}
 &V^{\rm L} = \frac{7a}{24\ell^2} \left \langle u^{\rm L}_{\rm A} \dot{u}^{\rm L}_{\rm B}-u^{\rm L}_{\rm B} \dot{u}^{\rm L}_{\rm A} \right \rangle
\nonumber \\
& - \frac{a\ell }{D^3}\left \langle 
u^{\rm R}_{\rm A}\dot{u}^{\rm R}_{\rm B}
- u^{\rm R}_{\rm B}\dot{u}^{\rm R}_{\rm A}
-u^{\rm L}_{\rm A}\dot{u}^{\rm R}_{\rm A} 
-u^{\rm L}_{\rm A}\dot{u}^{\rm R}_{\rm B}
+u^{\rm L}_{\rm B}\dot{u}^{\rm R}_{\rm A}
+u^{\rm L}_{\rm B}\dot{u}^{\rm R}_{\rm B}
\right \rangle,
\label{lvellr}\\
& V^{\rm R} = \frac{7a}{24\ell^2}\left \langle u^{\rm R}_{\rm A} \dot{u}^{\rm R}_{\rm B}-u^{\rm R}_{\rm B} \dot{u}^{\rm R}_{\rm A} \right \rangle
\nonumber \\
& - \frac{a\ell}{ D^3}\left \langle 
u^{\rm L}_{\rm A}\dot{u}^{\rm L}_{\rm B}
-u^{\rm L}_{\rm B}\dot{u}^{\rm L}_{\rm A}
-u^{\rm R}_{\rm A}\dot{u}^{\rm L}_{\rm A} 
-u^{\rm R}_{\rm A}\dot{u}^{\rm L}_{\rm B}
+u^{\rm R}_{\rm B}\dot{u}^{\rm L}_{\rm A}
+u^{\rm R}_{\rm B}\dot{u}^{\rm L}_{\rm B}
\right \rangle.
\label{rvellr}
\end{align}
Note that we have kept only up to second-order terms in $u^{\rm L,R}_{\rm A,B}$ as in 
Ref.~\cite{Najafi10}, meaning that we are also assuming the condition 
$u^{\rm L,R}_{\rm A,B}/\ell \ll \ell/D$.
The first terms on the right-hand side of the above equations represent the average swimming velocity 
of a single three-sphere swimmer, as previously obtained by Golestanian and Ajdari~\cite{Golestanian08}. 
These terms indicate that the average velocity of an isolated three-sphere swimmer is determined by 
the area enclosed by the orbit of the periodic motion in the configuration space.

The second terms on the right-hand side of Eqs.~(\ref{lvellr}) and (\ref{rvellr}) are due to the 
hydrodynamic interaction between the two swimmers.
These correction terms decay as $(\ell/D)^3$ with increasing distance because they result from force 
quadrupoles rather than force dipoles~\cite{Golestanian08}.
In fact, such a cubic dependence originates from the symmetry such that the motion of 
three-sphere swimmers is invariant under a combined time-reversal and parity transformation~\cite{Pooley07}.
The correction terms $\left \langle u^{\rm R}_{\rm A}\dot{u}^{\rm R}_{\rm B} 
- u^{\rm R}_{\rm B}\dot{u}^{\rm R}_{\rm A} \right \rangle$ in $V^{\rm L}$
and $\left \langle u^{\rm L}_{\rm A}\dot{u}^{\rm L}_{\rm B} 
-u^{\rm L}_{\rm B}\dot{u}^{\rm L}_{\rm A} \right \rangle$ in  $V^{\rm R}$ are both passive 
terms because they correspond to the swimming of only the second swimmer.
The other correction terms are due to the simultaneous motion of the two swimmers and 
hence are called active terms~\cite{Pooley07,Farzin12}.
We show later that only the active terms depend on the phase difference between the two swimmers.

\section{Two interacting elastic swimmers}
\label{sec:elastic}

In this section, we consider two interacting elastic three-sphere swimmers, as schematically shown in 
Fig.~\ref{fig:twoswimmerpic}, and calculate their average velocities.
We first assume that these two elastic swimmers have identical structures, whereas the structure 
of each swimmer can be either asymmetric or symmetric (as separately discussed in 
Sect.~\ref{sec:symmetric}).
For each swimmer, the two spring constants of harmonic springs A and B are denoted by $K_{\rm A}$ 
and $K_{\rm B}$, respectively.
Then the total energy of these two elastic swimmers is given by
\begin{equation}
\begin{split}
E& =\frac{K_{\rm A}}{2}\left(x_2-x_1-\ell^{\rm L}_{\rm A}\right)^2
+\frac{K_{\rm B}}{2}\left (x_3-x_2-\ell^{\rm L}_{\rm B}\right)^2
\\
&+\frac{K_{\rm A}}{2}\left(x_5-x_4-\ell^{\rm R}_{\rm A}\right)^2
+\frac{K_{\rm B}}{2}\left(x_6-x_5-\ell^{\rm R}_{\rm B}\right)^2.
\label{energy}
\end{split}
\end{equation}
In the above, $\ell^{\rm L}_{\rm A}$, $\ell^{\rm L}_{\rm B}$, $\ell^{\rm R}_{\rm A}$, and $\ell^{\rm R}_{\rm B}$
are the natural lengths of the respective harmonic springs and generally depend on time.
Hence, the six forces in Eq.~(\ref{ssokudo}) are given by  
\begin{equation}
f_i=-\frac{\partial E}{\partial x_i}.
\label{forcevel}
\end{equation}

For these two elastic swimmers, we assume that the four natural lengths of the springs undergo the following 
periodic changes in time~\cite{Farzin12}:
\begin{align}
\ell^{\rm L}_{\rm A}(t)&=\ell+d_{\rm A}\cos(\Omega t), 
\label{ellLA}
\\
\ell^{\rm L}_{\rm B}(t)&=\ell+d_{\rm B}\cos(\Omega t-\phi), 
\\
\ell^{\rm R}_{\rm A}(t)&=\ell+d_{\rm A}\cos(\Omega t-\Psi), 
\\
\ell^{\rm R}_{\rm B}(t)&=\ell+d_{\rm B}\cos(\Omega t-\phi-\Psi).
\label{ellRB}
\end{align}
Here, $\ell$ is the common average length as introduced in Eqs.~(\ref{uL}) and 
(\ref{uR}), $d_{\rm A}$ and $d_{\rm B}$ are the amplitudes of the oscillatory change, 
$\Omega$ is the common frequency, $\phi$ is the relative 
phase difference between the two springs within the swimmers, and $\Psi$ is the relative phase difference 
between the left and right swimmers. 
For each swimmer to move by itself, the time-reversal symmetry of the spring dynamics should be broken, 
i.e., $\phi \neq 0$.
In the absence of the hydrodynamic interaction between the two swimmers, they move in the same direction 
with the same velocity.
Although this assumption can be relaxed, the current situation already provides us with very rich dynamical 
behaviors when they interact hydrodynamically.
We also note that the frequency $\Omega$ can be different between the two swimmers, but such a study
is left as a future work.

It is convenient to introduce a characteristic time scale defined by~\cite{Yasuda17} 
\begin{equation}
\tau=\frac{6\pi\eta a}{K_{\rm A}}.
\label{tau}
\end{equation}
Then we use $\ell$ to scale all the relevant lengths and employ $\tau$  to scale the frequency, 
i.e., $\hat \Omega = \Omega \tau$.
By further defining the ratio between the two spring constants as  
$\lambda = K_{\rm B}/K_{\rm A}$, the coupled equations can be made dimensionless.
These equations can be solved in the frequency domain, and we further obtain 
$u^{\rm L,R}_{\rm A,B}$ in Eqs.~(\ref{kineleft}) and (\ref{kineright}) after an inverse 
Fourier transform~\cite{Yasuda17}.
Since their explicit expressions are somewhat lengthy, we give them in 
Appendix~\ref{app:displacements}.
Finally, using Eqs.~(\ref{lvellr}) and (\ref{rvellr}), we calculate the average velocities 
$V^{\rm L}$ and $V^{\rm R}$ of the two elastic swimmers.
Their full expressions are given in Appendix~\ref{app:velocities}.

\begin{figure}[tbh]
\centering
\includegraphics[scale=0.3]{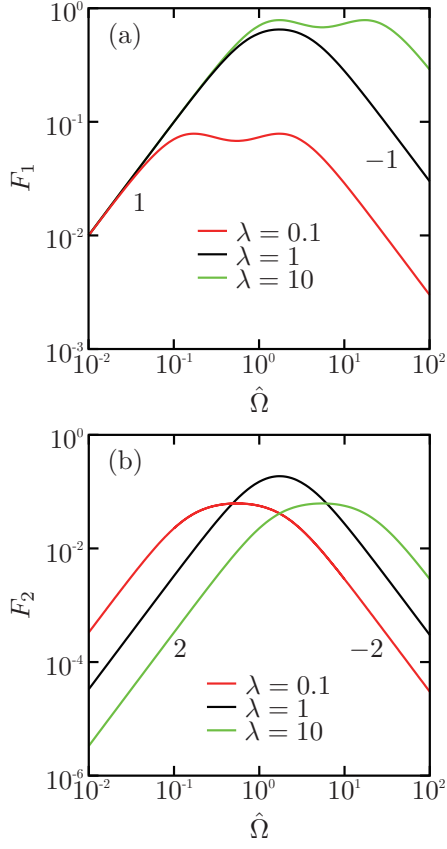}
\caption{(Color online)
Plots of the scaling functions (a) $F_1(\hat \Omega; \lambda)$ and (b) 
$F_2(\hat \Omega; \lambda)$ defined in Eqs.~(\ref{eq:f1}) and (\ref{eq:f2}), respectively, 
as functions of $\hat \Omega = \Omega \tau$ for $\lambda= K_{\rm B}/K_{\rm A}=0.1, 1,$ and $10$.
The numbers indicate the slope, representing the exponent of the power-law behaviors.}
\label{fig2}
\end{figure}

\begin{figure}[tbh]
\centering
\includegraphics[scale=0.3]{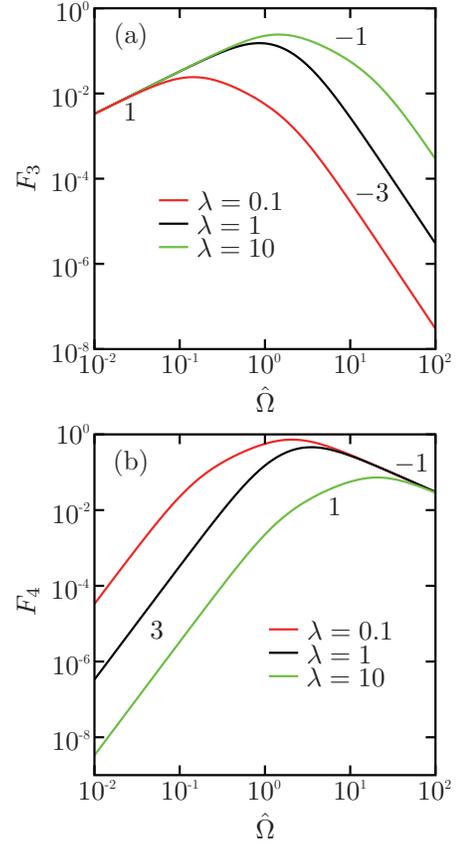}
\caption{(Color online)
Plots of the scaling functions (a) $F_3(\hat \Omega; \lambda)$ and (b) 
$F_4(\hat \Omega; \lambda)$ defined in Eqs.~(\ref{eq:f3}) and (\ref{eq:f4}), respectively, 
as functions of $\hat \Omega = \Omega \tau$ for $\lambda= K_{\rm B}/K_{\rm A}=0.1, 1,$ and $10$.
The numbers indicate the slope, representing the exponent of the power-law behaviors.}
\label{fig3}
\end{figure}

Instead, we show here the mean and the difference between the two average velocities 
$V^{\rm L}$ and $V^{\rm R}$.
The former is given by 
\begin{align}
\frac{V^{\rm R}+V^{\rm L}}{2}=V_0\left[1-\frac{48\ell^3}{7D^3}\sin^2(\Psi/2)\right],
\label{eq:ave}
\end{align}
where the average velocity of a single elastic swimmer was obtained before as~\cite{Yasuda17}
\begin{align}
V_0&=\frac{7d_{\rm A}d_{\rm B}a}{24\ell^2\tau}F_1(\hat{\Omega};\lambda)\sin\phi
\nonumber \\
&+\frac{7(1-\lambda)d_{\rm A}d_{\rm B}a}{12\ell^2\tau}F_2(\hat{\Omega};\lambda)\cos\phi
\nonumber \\
&+\frac{7(d^2_{\rm A}-d^2_{\rm B}\lambda)a}{24\ell^2\tau}F_2(\hat{\Omega};\lambda).
\label{eq:v1g}
\end{align}
On the other hand, the velocity difference between the two swimmers is given by  
\begin{align}
V^{\rm R}-V^{\rm L}& =\frac{a\ell}{D^3\tau}
\left[ 2 d_{\rm A}d_{\rm B}(1+\lambda)F_2(\hat{\Omega};\lambda)\sin\phi
\right. 
\nonumber \\
& -3(d_{\rm A}^2-d_{\rm B}^2) F_3(\hat{\Omega};\lambda)
\nonumber \\
& \left. - (d_{\rm A}^2-d_{\rm B}^2\lambda^2) F_4(\hat{\Omega};\lambda) \right] \sin\Psi.
\label{eq:deltv}
\end{align}
In the above equations, we have introduced four scaling functions defined by 
\begin{align}
F_1(\hat{\Omega};\lambda)& =\frac{3\lambda\hat{\Omega}(3\lambda+\hat{\Omega}^2)}{9\lambda^2+2(2+\lambda+2\lambda^2)\hat{\Omega}^2+\hat{\Omega}^4},
\label{eq:f1}
\\
F_2(\hat{\Omega};\lambda)& =\frac{3\lambda\hat{\Omega}^2}{9\lambda^2+2(2+\lambda+2\lambda^2)\hat{\Omega}^2+\hat{\Omega}^4},
\label{eq:f2}
\\
F_3(\hat{\Omega};\lambda)& =\frac{3\lambda^2\hat{\Omega}}{9\lambda^2+2(2+\lambda+2\lambda^2)\hat{\Omega}^2+\hat{\Omega}^4},
\label{eq:f3}
\\
F_4(\hat{\Omega};\lambda)& =\frac{3\hat{\Omega}^3}{9\lambda^2+2(2+\lambda+2\lambda^2)\hat{\Omega}^2+\hat{\Omega}^4}.
\label{eq:f4}
\end{align}
These are the main results of this paper.

Equation~(\ref{eq:ave}) indicates that, owing to the hydrodynamic interaction 
between the two elastic swimmers, the mean velocity is always smaller than $V_0$ 
irrespective of the relative phase difference $\Psi$. 
Using the formula $\sin^2(\Psi/2)=(1-\cos \Psi)/2$ in Eq.~(\ref{eq:ave}), we point out that the 
$\Psi$-independent contribution to the correction is due to the passive terms in Eqs.~(\ref{lvellr}) 
and (\ref{rvellr}), whereas the $\Psi$-dependent contribution comes from the active terms.
The correction to $V_0$ vanishes only when $\Psi=0$, and the mean velocity is minimized when $\Psi = \pi$.
In contrast, the velocity difference in Eq.~(\ref{eq:deltv}) can be either positive or negative depending 
on the conditions. 
Obviously, we have $V^{\rm R}=V^{\rm L}$ when $\Psi=0$. 
This is reasonable because the two swimmers should move with the same velocity when the relative phase 
difference vanishes. 
A more detailed discussion concerning the velocity difference will be given in the next section for symmetric elastic
swimmers.

In Fig.~\ref{fig2}, we plot the scaling functions $F_1$ and $F_2$ as functions of $\hat \Omega$ for 
$\lambda=0.1$, $1$, and $10$~\cite{Yasuda17}. 
Note, however, that the cases of $\lambda=0.1$ and $10$ are essentially equivalent because we can always 
exchange springs A and B, whereas we have defined the relaxation time $\tau$ through $K_{\rm A}$
as in Eq.~(\ref{tau}).
As shown in Eq.~(\ref{eq:v1g}) and previously discussed in Ref.~\cite{Yasuda17}, the frequency dependence of the 
average velocity $V_0$ for an isolated elastic swimmer is essentially determined by 
$F_1(\hat{\Omega};\lambda)$ and $F_2(\hat{\Omega};\lambda)$. 
Notice that $F_1 \sim \Omega$ and $F_2 \sim \Omega^2$ for $\hat \Omega \ll 1$, whereas 
$F_1 \sim \Omega^{-1}$ and $F_2 \sim \Omega^{-2}$ for $\hat \Omega \gg 1$.
Hence the average velocity increases for $\hat \Omega \ll 1$, whereas it decreases for 
$\hat \Omega \gg 1$ when the frequency is increased~\cite{Yasuda17}.
To ensure the validity of our elastic swimmer model, we assume here that 
a low-Reynolds-number flow field is justified even in the high-frequency regime 
$\hat \Omega \gg 1$.

In Fig.~\ref{fig3}, we plot the scaling functions $F_3$ and $F_4$ as functions of $\hat \Omega$ for 
$\lambda=0.1$, $1$, and $10$. 
As shown in Eq.~(\ref{eq:deltv}), the scaling functions $F_3(\hat{\Omega};\lambda)$ and $F_4(\hat{\Omega};\lambda)$ 
as well as $F_2(\hat{\Omega};\lambda)$ characterize the frequency dependence of the hydrodynamic interaction 
between two elastic swimmers. 
Here, we have $F_3 \sim \Omega$ and $F_4 \sim \Omega^3$ for $\hat \Omega \ll 1$, whereas 
$F_3 \sim \Omega^{-3}$ and $F_4 \sim \Omega^{-1}$ for $\hat \Omega \gg 1$.
When the swimmers are asymmetric such as when $\lambda=0.1$ and $10$, on the other hand, there are 
intermediate regions where the scaling functions behave as $F_3 \sim \Omega^{-1}$ and $F_4 \sim \Omega$.
Note that the velocity difference also decreases in the high-frequency regime.

\section{Two symmetric elastic swimmers}
\label{sec:symmetric}

\begin{figure}[tbh]
\centering
\includegraphics[scale=0.3]{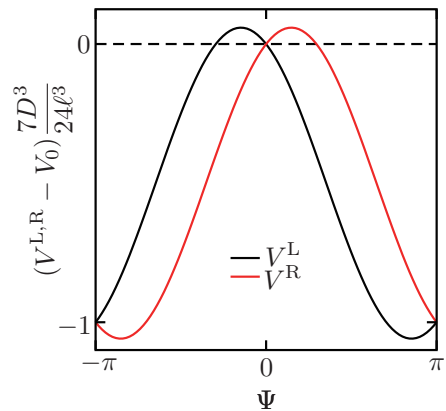}
\caption{(Color online)
Average velocities $V^{\rm L}$ (black) and $V^{\rm R}$ (red) of two symmetric 
elastic swimmers with respect to $V_0$ as a function of the relative phase difference $\Psi$ between 
them when $\hat{\Omega}=1$.
See Eqs.~(\ref{lswimvel}) and (\ref{rswimvel}).}
\label{velg}
\end{figure}

Having discussed the case of two general (asymmetric) elastic swimmers, we now discuss the 
case when both elastic swimmers have symmetric structures, i.e.,
$d_{\rm A}=d_{\rm B}=d$ and $K_{\rm A}=K_{\rm B}$ (or $\lambda=1$).
In this case, the two average velocities can be simply written as 
\begin{align}
V^{\rm L}
&=V_0\left[ 1-\frac{48\ell^3}{7D^3}\left(\sin^2(\Psi/2)+\frac{\hat{\Omega}}{3+\hat{\Omega}^2}\sin\Psi\right)\right],
\label{lswimvel}\\
V^{\rm R}
&=V_0\left[1-\frac{48\ell^3}{7D^3}\left(\sin^2(\Psi/2)-\frac{\hat{\Omega}}{3+\hat{\Omega}^2}\sin\Psi\right)\right],
\label{rswimvel}
\end{align}
where the average velocity of a single elastic swimmer now becomes~\cite{Yasuda17} 
\begin{equation}
V_0=\frac{7d^2a}{24\ell^2\tau}\frac{3\hat{\Omega}(3+\hat{\Omega}^2)}{9+10\hat{\Omega}^2+\hat{\Omega}^4}\sin\phi.
\label{eq:v1}
\end{equation}
In Eqs.~(\ref{lswimvel}) and (\ref{rswimvel}), the $\Psi$-independent terms of 
$\sin^2(\Psi/2)=(1-\cos \Psi)/2$ correspond to the passive terms as before.
In Fig.~\ref{velg}, we plot the $\Psi$-dependences of $V^{\rm L}-V_0$ and $V^{\rm R}-V_0$ when 
$\hat{\Omega}=1$.
We see that both $V^{\rm L}$ and $V^{\rm R}$ can be larger than $V_0$ for certain ranges of $\Psi$.
For $\hat{\Omega}=1$, as shown in Fig.~\ref{velg}, we have 
$V^{\rm L}>V_0$ for $-0.927< \Psi <0$ and $V^{\rm R}>V_0$ for $0< \Psi < 0.927$.
However, as we have already explained with Eq.~(\ref{eq:ave}) for the general asymmetric case, the mean 
of $V^{\rm L}$ and $V^{\rm R}$ is always smaller than $V_0$.

Furthermore, the velocity difference is now given by 
\begin{equation}
V^{\rm R}-V^{\rm L}= \frac{4 d^2 a\ell }{D^3\tau}  
\frac{3 \hat{\Omega}^2}{9+10\hat{\Omega}^2+\hat{\Omega}^4}\sin\phi \sin\Psi.
\label{difvel}
\end{equation}
This is an interesting result because, for $0 < \phi < \pi$ and hence $V_0>0$, we have $V^{\rm L}< V^{\rm R}$ 
for $0<\Psi <\pi$ or $V^{\rm L}> V^{\rm R}$ for $-\pi<\Psi <0$. 
In the former case, the interaction between the two swimmers is repulsive and the distance between 
them increases as they move, i.e., an unbound state.
In the latter case, on the other hand, the interaction is attractive and they form a moving hydrodynamic 
bound state.

It is worthwhile noting that, in the case of  $\hat{\Omega} \ll 1$, 
the average velocity in Eq.~(\ref{eq:v1}) behaves as $V_0 \sim \hat{\Omega}$,  
whereas the velocity difference in Eq.~(\ref{difvel}) scales as 
$V^{\rm R}-V^{\rm L} \sim \hat{\Omega}^2$
for two symmetric swimmers.
Such a difference arises from the presence of the active terms in Eqs.~(\ref{lvellr}) and (\ref{rvellr})
[or the last $\hat{\Omega}$-dependent terms in Eqs.~(\ref{lswimvel}) and (\ref{rswimvel})]
owing to the simultaneous motion of the two swimmers.
According to the above frequency dependences, the velocity difference $V^{\rm R}-V^{\rm L}$
(for finite $D$) becomes much smaller than $V_0$ in the limit of $\hat{\Omega} \rightarrow 0$, 
and the two velocities turn out to be identical, as shown later in Eq.~(\ref{rgolswimvel}).
A similar argument holds also for $\hat{\Omega} \gg 1$ because we have 
$V_0 \sim \hat{\Omega}^{-1}$ and $V^{\rm R}-V^{\rm L} \sim \hat{\Omega}^{-2}$, 
the latter being the higher-order active contribution.

\section{Limit of two NG swimmers}
\label{sec:NG}

The interaction between two asymmetric NG swimmers can be recovered simply by 
taking the limit of $\hat \Omega =\Omega \tau \rightarrow 0$.
This is because the spring constants $K_{\rm A}$ and $K_{\rm B}$ are infinitely large and the 
characteristic time scale $\tau=6\pi\eta a/K_{\rm A}$ is infinitely small for NG swimmers.
In this limit, the two average velocities defined by 
$v^{\rm L,R}=V^{\rm L,R}(\hat{\Omega} \rightarrow 0)$ become 
\begin{align}
v^{\rm L} & = v_0 - \frac{a\ell \Omega}{2 D^3}\left[ 4d_{\rm A}d_{\rm B}\sin^2(\Psi/2)\sin\phi
\right.
\nonumber \\
& \left.  - (d_{\rm A}^2-d_{\rm B}^2)\sin\Psi \right],
\label{goll}
\end{align}
\begin{align}
v^{\rm R}& =v_0  - \frac{a\ell \Omega}{2 D^3}\left[ 4d_{\rm A}d_{\rm B}\sin^2(\Psi/2)\sin\phi
\right.
\nonumber \\
& \left.  + (d_{\rm A}^2-d_{\rm B}^2)\sin\Psi \right],
\label{golr}
\end{align}
where the average velocity of a single NG swimmer is~\cite{Golestanian08}
\begin{align}
v_0=\frac{7d_{\rm A}d_{\rm B} a\Omega}{24\ell^2}\sin\phi.
\label{gol1}
\end{align}
Hence, the mean of $v^{\rm L}$ and $v^{\rm R}$ is again given by Eq.~(\ref{eq:ave})
in which $V_0$ is replaced by $v_0$.
As mentioned in the previous section, both $v^{\rm L}$ and $v^{\rm R}$ are proportional 
to $\Omega$.
The velocity difference, on the other hand, becomes 
\begin{align}
v^{\rm R}-v^{\rm L}=-\frac{a\ell \Omega}{D^3}(d^2_{\rm A}-d^2_{\rm B})\sin{\Psi}.
\label{golans}
\end{align}
This result indicates that the velocity difference depends not only on $\Psi$ but 
also on the relative magnitude between $d_{\rm A}$ and $d_{\rm B}$ for asymmetric NG swimmers.

For symmetric NG swimmers, i.e., $d_{\rm A}=d_{\rm B}$, $v^{\rm L}$ and $v^{\rm R}$ are 
identical and are given by   
\begin{align}
v^{\rm L} =v^{\rm R}=v_0 \left[1-\frac{48\ell^3}{7D^3}\sin^2(\Psi/2)\right].
\label{rgolswimvel}
\end{align}
This result means that the average velocities $v^{\rm L}$ and $v^{\rm R}$ of the two symmetric 
NG swimmers are always smaller than that of an isolated NG swimmer, 
i.e., $v^{\rm L,R}< v_0$.
Hence, the possibility of $V^{\rm L,R}> V_0$ under certain conditions, as shown 
in Eqs.~(\ref{lswimvel}) and (\ref{rswimvel}), is a unique feature of two elastic swimmers.
Since $v^{\rm L} =v^{\rm R}$ for two symmetric NG swimmers, the distance between them 
remains constant, which is in contrast to the case of two symmetric elastic swimmers
[see Eq.~(\ref{difvel})].
Such a difference arises from the internal relaxation dynamics of the spheres in elastic swimmers,
leading to asymmetric motion of the two springs in each swimmer.

\section{Summary and discussion}
\label{sec:summary}

We have investigated the hydrodynamic interaction between two elastic swimmers
consisting of three spheres and two harmonic springs. 
In this model, the natural length of each spring is assumed to undergo a prescribed cyclic 
change in time, reflecting the internal states of an elastic swimmer.
For two interacting three-sphere microswimmers, we first obtained their average velocities in 
terms of the distance $D$ between them [see Eqs.~(\ref{lvellr}) and (\ref{rvellr})].
Using these expressions, we further obtained the explicit forms of the average velocities 
of two identical elastic swimmers. 
The mean of the two average velocities was shown to be always smaller than that of a single elastic 
swimmer [see Eq.~(\ref{eq:ave})]. 
On the other hand, the velocity difference depends on the relative phase difference $\Psi$ 
between the two elastic swimmers [see Eqs.~(\ref{eq:deltv}) and (\ref{difvel})].
As a result, the swimming state of two elastic swimmers can be either bound or unbound 
depending on the relative phase difference.

In this paper, the hydrodynamic interaction was considered only between two three-sphere 
microswimmers, although there are several other model swimmers. 
For example, two rigid helices neither attract nor repel each other when they are rotating 
with zero phase difference~\cite{Kim04}, 
two puller-type squirmers undergo a significant change in their 
orientations after an encounter~\cite{Ishikawa06}, 
and two spherical swimmers with spatially confined circular trajectories cause either attractive 
or repulsive interaction~\cite{Michelin10}.
Using the Quadroar model, Mirzakhanloo \textit{et al.}\ showed that two swimmers, which generate 
flow fields mimicking that of \textit{Chlamydomonas reinhardtii}, exhibit very 
rich behaviors~\cite{Mirzakhanloo18}.
The three-sphere swimmer model in one-dimensional space is especially suitable for analytical analysis 
because it is sufficient to consider only the translational motion, and the tensorial structure 
of the fluid motion can be neglected.

In our work, we have assumed that the two elastic three-sphere swimmers are 
confined in one-dimensional space and moving in the same direction.
For two NG swimmers, on the other hand, it was shown before that the interaction between 
them depends on their relative orientation~\cite{Pooley07,Farzin12}. 
The main reason that we have investigated only the one-dimensional case is that our 
primary interest is to analytically obtain the frequency dependence of the hydrodynamic 
interaction between two elastic swimmers, which was not studied before. 
Another motivation to restrict our study to one-dimensional space is to clarify how the 
correlation between a three-sphere swimmer and a two-sphere system, as reported 
in Ref.~\cite{Najafi10}, can be generalized for two three-sphere swimmers [see 
Eqs.~(\ref{lvellr}) and (\ref{rvellr})].
The future study of the hydrodynamic interaction between two elastic swimmers having 
different orientations would require a numerical treatment. 
For instance, the oscillatory motion reported in Refs.~\cite{Pooley07} and \cite{Farzin12}, 
would be observable only when the space dimension is higher than one.

We have shown analytically that even the interaction between two elastic microswimmers can be 
complicated, depending on the relative displacement, structure, and phase difference.
Nevertheless, it is possible and straightforward to increase the number of interacting swimmers 
as long as the assumption of low-Reynolds-number hydrodynamics is valid and the swimmers 
are confined in one-dimensional space.
We believe that the present analysis of the hydrodynamic interaction between two swimmers 
will be useful in studying the collective behavior of a large number of self-propelled 
microswimmers immersed in a viscous fluid~\cite{Stenhammar17,Filella18}.

\acknowledgements

We thank S.\ Al-Izzi, H.-Y.\ Chen, Y.\ Hosaka, T.\ Kato, H.\ Kitahata, Y.\ Koyano, and R.\ Okamoto 
for fruitful discussions and helpful suggestions.
K.Y.\ acknowledges support by a Grant-in-Aid for JSPS Fellows (Grant No. 18J21231) from the Japan 
Society for the Promotion of Science (JSPS). 
S.K. acknowledges support by a Grant-in-Aid for Scientific Research (C) (Grant No. 18K03567) from 
the JSPS.

\appendix
\section{Displacements $u^{\rm L}_{\rm A}$, $u^{\rm L}_{\rm B}$, $u^{\rm R}_{\rm A}$, $u^{\rm R}_{\rm B}$}
\label{app:displacements}

The four displacements 
$u^{\rm L}_{\rm A}$, $u^{\rm L}_{\rm B}$, $u^{\rm R}_{\rm A}$, and $u^{\rm R}_{\rm B}$
of two interacting elastic swimmers are given as follows:
\begin{align}
\label{eq:ualam}
u^{\rm L}_{\rm A}=& \frac{1}{9\lambda^2+2(2+\lambda+2\lambda^2)\hat{\Omega}^2+\hat{\Omega}^4}\nonumber\\
&\times\Bigl\{[9\lambda^2+(4+\lambda)\hat{\Omega}^2] d_{\rm A}\cos(\Omega t)\nonumber\\
&+2(3\lambda^2+\hat{\Omega}^2)\hat{\Omega} d_{\rm A}\sin(\Omega t)\nonumber\\
&-2\lambda(1+\lambda)\hat{\Omega}^2d_{\rm B}\cos(\Omega t-\phi)\nonumber\\
&-\lambda(-3\lambda+\hat{\Omega}^2)\hat{\Omega}d_{\rm B}\sin(\Omega t-\phi)\Bigr\},\\
\label{eq:ublam}
u^{\rm L}_{\rm B}=&\frac{1}{9\lambda^2+2(2+\lambda+2\lambda^2)\hat{\Omega}^2+\hat{\Omega}^4}\nonumber\\
&\times\Bigl\{ -2(1+\lambda)\hat{\Omega}^2d_{\rm A}\cos(\Omega t)\nonumber\\
&+(3\lambda-\hat{\Omega}^2)\hat{\Omega}d_{\rm A}\sin(\Omega t)\nonumber\\
&+\lambda [9\lambda+(1+4\lambda)\hat{\Omega}^2]d_{\rm B}\cos(\Omega t-\phi)\nonumber\\
&+2\lambda(3+\hat{\Omega}^2)\hat{\Omega}d_{\rm B}\sin(\Omega t-\phi)\Bigr\},\\
\label{eq:uaplam}
u^{\rm R}_{\rm A}=&\frac{1}{9\lambda^2+2(2+\lambda+2\lambda^2)\hat{\Omega}^2+\hat{\Omega}^4}\nonumber\\
&\times\Bigl\{ [9\lambda^2+(4+\lambda)\hat{\Omega}^2]d_{\rm A}\cos(\Omega t-\Psi)\nonumber\\
&+2(3\lambda^2+\hat{\Omega}^2)\hat{\Omega}d_{\rm A}\sin(\Omega t-\Psi)\nonumber\\
&-2\lambda(1+\lambda)\hat{\Omega}^2d_{\rm B}\cos(\Omega t-\phi-\Psi)\nonumber\\
&-\lambda(-3\lambda+\hat{\Omega}^2)\hat{\Omega}d_{\rm B}\sin(\Omega t-\phi-\Psi)\Bigr\},\\
\label{eq:ubplam}
u^{\rm R}_{\rm B}=&\frac{1}{9\lambda^2+2(2+\lambda+2\lambda^2)\hat{\Omega}^2+\hat{\Omega}^4}\nonumber\\
&\times\Bigl\{-2(1+\lambda)\hat{\Omega}^2d_{\rm A}\cos(\Omega t-\Psi)\nonumber\\
&+(3\lambda-\hat{\Omega}^2)\hat{\Omega}d_{\rm A}\sin(\Omega t-\Psi)\nonumber\\
&+\lambda [9\lambda+(1+4\lambda)\hat{\Omega}^2 ]d_{\rm B}\cos(\Omega t-\phi-\Psi)\nonumber\\
&+2\lambda(3+\hat{\Omega}^2)\hat{\Omega}d_{\rm B}\sin(\Omega t-\phi-\Psi)\Bigr\}.
\end{align}

\section{Average velocities $V^{\rm L}$ and $V^{\rm R}$}
\label{app:velocities}

The average velocities $V^{\rm L}$ and $V^{\rm R}$ of two interacting elastic swimmers are 
given as follows:
\begin{align}
V^{\rm L}  & = V_0-\frac{a\ell}{D^3\tau}\Bigl[
2d_{\rm A}d_{\rm B}\sin^2(\Psi/2)F_1(\hat{\Omega};\lambda)\sin\phi
\nonumber \\
&+d_{\rm A}d_{\rm B}(1+\lambda)F_2(\hat{\Omega};\lambda)\sin\Psi\sin\phi
\nonumber \\
&+4d_{\rm A}d_{\rm B}(1-\lambda)F_2(\hat{\Omega};\lambda)\sin^2(\Psi/2)\cos\phi
\nonumber \\
&+2(d_{\rm A}^2-d_{\rm B}^2\lambda)F_2(\hat{\Omega};\lambda)\sin^2(\Psi/2)
\nonumber \\
&-\frac{1}{2}[3(d_{\rm A}^2-d_{\rm B}^2)F_3(\hat{\Omega};\lambda)
\nonumber \\
&+(d_{\rm A}^2-d_{\rm B}^2\lambda^2)F_4(\hat{\Omega};\lambda)]\sin\Psi
\Bigr],
\label{eq:lamres}
\end{align}

\begin{align}
V^{\rm R} & = V_0-\frac{a\ell}{D^3\tau}\Bigl[
2d_{\rm A}d_{\rm B}\sin^2(\Psi/2)F_1(\hat{\Omega};\lambda)\sin\phi
\nonumber \\
&-d_{\rm A}d_{\rm B}(1+\lambda)F_2(\hat{\Omega};\lambda)\sin\Psi\sin\phi
\nonumber \\
&+4d_{\rm A}d_{\rm B}(1-\lambda)F_2(\hat{\Omega};\lambda)\sin^2(\Psi/2)\cos\phi
\nonumber \\
&+2(d_{\rm A}^2-d_{\rm B}^2\lambda)F_2(\hat{\Omega};\lambda)\sin^2(\Psi/2)
\nonumber \\
&+\frac{1}{2}[3(d_{\rm A}^2-d_{\rm B}^2)F_3(\hat{\Omega};\lambda)
\nonumber \\
&+(d_{\rm A}^2-d_{\rm B}^2\lambda^2)F_4(\hat{\Omega};\lambda)]\sin\Psi
\Bigr],
\label{eq:ramres}
\end{align}
where $V_0$ is given by Eq.~(\ref{eq:v1g}) and the four scaling functions are given by 
Eqs.~(\ref{eq:f1})--(\ref{eq:f4}).


\end{document}